\documentclass[a4paper]{article}

\usepackage{amsmath,amssymb,amsfonts,amsthm}
\usepackage{mathrsfs}
\usepackage{hyperref}
\usepackage{enumitem}
\usepackage{bbm}
\usepackage[affil-it]{authblk}

\newtheorem{theorem}{Theorem}
\newtheorem{lemma}{Lemma}
\newtheorem{proposition}{Proposition}

\newtheorem{definition}{Definition}
\newtheorem{example}{Example}
\newtheorem{remark}{Remark}
\newtheorem{assumption}{Assumption}

\title{On Sybil Proofness in Competitive Combinatorial Exchanges}

\author{Abhimanyu Nag}
\affil{
Department of Mathematical and Statistical Sciences \\
University of Alberta, Canada}

\date{}

\begin{document}

\maketitle

\begin{abstract}
We study Sybil manipulation in BRACE, a competitive equilibrium mechanism for combinatorial exchanges, by treating identity creation as a finite perturbation of the empirical distribution of reported types. Under standard regularity assumptions on the excess demand map and smoothness on principal utilities, we obtain explicit linear bounds on price and welfare deviations induced by bounded Sybil invasion. Using these bounds, we prove a sharp contrast: strategyproofness in the large holds if and only if each principal’s share of identities vanishes whereas any principal with a persistent positive share can construct deviations yielding strictly positive limiting gains. We further show that the feasibility of BRACE fails in the event of an unbounded population of Sybils and provide a precise cost threshold which ensures disincentivization of such attacks in large markets.
\end{abstract}

\section{Introduction}

Combinatorial exchanges \cite{budish2011combinatorial} provide a general framework for allocating indivisible goods without transferable utility. A recent breakthrough in this literature is the \emph{Budget-Relaxed Approximate Competitive Equilibrium} (BRACE) mechanism of Jantschgi, Teytelboym and Nguyen \cite{jantschgi2025competitive}. By introducing a random budget relaxation, BRACE smooths the discontinuities inherent to discrete demand and restores competitive equilibrium (CE) (Hylland and Zeckhauser \cite{hylland1979efficient}). The mechanism guarantees approximate feasibility, individual rationality, ordinal efficiency, justified envy-freeness (see Yilmaz \cite{yilmaz2010probabilistic}) and \emph{strategyproofness in the large} (SP-L) (Azevedo and Budish \cite{azevedo2019strategy}). Under BRACE, the influence of any single reporting identity vanishes as the market grows.

However, BRACE evaluates reports at the level of \emph{identities} (i.e a distinct strategic agent), not \emph{principals} (who may control and coordinate the behavior of multiple identities)\footnote{A detailed exposition about the definition of principals and identities can be found in the later sections}. In decentralized environments such as blockchain blockspace auctions, sequencer markets, validator rotations or other privacy preserving financial applications, identities are cheaply created and controlled by a smaller number of underlying principals \cite{saleh2021blockchain}. This discrepancy raises a fundamental question: Do the structural guarantees of BRACE survive when principals can create arbitrarily many identities? This strategic manipulation is termed as a Sybil attack \cite{douceur2002Sybil}, where a principal artificially inflates its presence by generating many fake identities. Despite a growing body of work and the centrality of such attacks in decentralized systems (see \cite{platt2023Sybil}), no prior work studies Sybil behaviour in competitive equilibrium based combinatorial exchanges.

\medskip

\noindent\textbf{Contributions.}
We developed a formal framework for analysing Sybil behaviour in competitive
combinatorial exchanges implemented through BRACE. Treating Sybil
creation as a perturbation of the empirical type distribution and under
assumptions on local regularity of excess demand, price uniqueness, and
Lipschitz utilities, we established a linear price-stability bound (Theorem \ref{thm:local-price}) which implies that Sybil attacks of bounded magnitude in the neighborhood would induce only bounded
price deviations. Combined with Lipschitz utilities, these bounds yield
correspondingly \emph{linear} welfare loss for every principal (Proposition \ref{prop:welfare}). Further, we demonstrated that BRACE’s fairness guarantees do not
aggregate across identities: justified envy-freeness may hold between identities while fairness fails amongst principals (Proposition \ref{prop:jef}).

Interestingly, we also prove that a principal can
affect the empirical distribution (of demand) by at most its identity share and BRACE is SP-L at the principal level \emph{if and only if} every
principal’s share of identities as a fraction of the total set of identities (denoted $s_{p,n}$) satisfies $\max_p s_{p,n} \to 0$. However, when a
principal controls a positive fraction of identities, its price impact does not
vanish and profitable coordinated misreports become possible even though
identity-level SP-L remains intact (Theorem \ref{prop:small-principals}).

One of the most important results showed that BRACE cannot exist under unbounded Sybil
mass if a principal's share tends to one and expected demand eventually violates even the basic requirements of equilibrium, precluding any sequence of $\delta$-BRACE
equilibria (Proposition \ref{prop:unbounded-Sybils}). Finally, by conditioning on the ``bad region'' where regularity
assumptions fail, we derived expected utility bounds showing that Sybil
sensitivity remains confined to whenever the probability of entering this region
vanishes (Lemma \ref{lem:tail-utility-bound}). Sufficient ground has been set up for future work that will deal with ``bad region'' more rigorously and provide better bounds.

On the incentive side, we combined identity-level SP-L gains with price-impact gains to obtain an explicit design inequality for deterrence of Sybil attacks which characterizes the minimal system-level cost required to eliminate
profitable Sybil strategies in sufficiently large economies (Proposition \ref{prop:design-inequality}).

\section{Background and Related Work}

\subsection{Combinatorial exchange and BRACE}
CE methods for discrete allocation originate in the work of Hylland and Zeckhauser \cite{hylland1979efficient}, Varian \cite{varian1973equity}, Budish \cite{budish2011combinatorial} and the pseudo market literature (See \cite{echenique2021constrained} for a detailed background discussion of equilibrium arguments in pseudo markets). These mechanisms typically rely on randomization or approximate feasibility to restore equilibrium existence in non-convex environments. Also see \cite{mittelmann2021general} for a complementary computational perspective on combinatorial exchanges. The BRACE mechanism of Jantschgi, Teytelboym and Nguyen \cite{jantschgi2025competitive} introduces \emph{random budget relaxation}, which smooths discontinuities in demand and enables a fixed-point argument establishing existence of $\delta$-BRACE equilibria for every $\delta>0$. BRACE inherits important welfare properties including individual rationality, ordinal efficiency, justified envy-freeness and ex-post realizability of random allocations. Moreover, BRACE is SP-L at the identity level with misreporting gains at a decay rate $O(n^{-1/2+\varepsilon})$ \cite{azevedo2019strategy}. Along with the definition and the aforementioned decay rate of SP-L, we also borrow $K(\epsilon)$ from Azevedo and Budish \cite{azevedo2019strategy} to derive our costs for continued Sybil attacks. Further we direct our readers to the original paper by Jantschgi et al. \cite{jantschgi2025competitive} for a detailed and comprehensive background about the literature on BRACE, combinatorial exchanges and incentive analysis techniques.

\subsection{Sybil Attacks}

In classical mechanism design, each reported identity is treated as a distinct
strategic agent and incentive and fairness guarantees are defined at the level of
these identities. In decentralized digital systems this assumption fails: public-key
identities are costless to generate, and a single principal may create arbitrarily
many identities at negligible cost. This phenomenon, first formalised by
Douceur’s Sybil attack model \cite{douceur2002Sybil}, is now recognised as a
central concern across peer-to-peer networks \cite{cai2014containing}, blockchains \cite{eisenbarth2022ethereum}
and cryptoeconomic security \cite{brekke2021hacker}. Yet the implications of identity replication have not
been analysed in competitive-equilibrium combinatorial exchanges.

Although there is a growing literature on Sybil proofness mechanisms
(e.g., \cite{pan2024Sybil,mazorra2023cost}), the effect of identity replication on competitive combinatorial exchanges allocation remains unexplored. The BRACE mechanism was
designed to recover equilibrium guarantees in markets with indivisible goods
and ordinal preferences, but its analysis implicitly assumes that each identity is a
genuine economic participant. Since BRACE relies on aggregate distributions in large populations, Sybil creation naturally raises the question of
how sensitive its allocations and prices are to adversarial perturbations of these
distributions.

These considerations motivate the central objective of this paper: to provide a
systematic analysis of Sybil creation in competitive combinatorial exchanges and
to determine which structural guarantees of BRACE survive when principals may
control multiple identities. Our approach draws on classical perturbation theory
for competitive equilibria, where stability is governed by the regularity of excess
demand \cite{iliopoulos1975functional}.

The paper progresses as follows: Section~\ref{sec:brace} reviews the BRACE allocation rule and its key theoretical
properties. Section~\ref{sec:Sybil} develops the core analysis of Sybil behaviour
in BRACE, establishing the main stability, welfare, and incentive results.
Section~\ref{sec:incentive} examines the cost structures required to deter Sybil
manipulation and ensure incentive compatibility. Section~\ref{conclusion}
concludes and outlines directions for future research.

\section{Formal Model of BRACE in Combinatorial Exchanges} \label{sec:brace}

In what follows, we provide a detailed formulation of the BRACE mechanism for competitive combinatorial exchanges, remaining faithful to the modelling conventions and assumptions of \cite{jantschgi2025competitive}.
While our presentation is self-contained, the reader is encouraged to consult the original BRACE paper for a full account of the mechanism’s derivation and foundational results. 

There is a finite set of \emph{identities} (agents) $N := \{1,\dots,n\}$ and a finite set of indivisible goods $M := \{1,\dots,m\}$.

Each good \(j\in M\) has an integer capacity \(c_j \in \mathbb{N}_{>0}\).  Let $c := (c_1,\dots,c_m) \in \mathbb{N}^m_{>0}$
denote the capacity vector.

A \emph{bundle} is an integral vector \(\textbf{x} \in \mathbb{N}^m\).  The number of units of good \(j\) in bundle \(x\) is denoted \(x_j\).  For each identity \(i\in N\), let $\Psi_i \subseteq \mathbb{N}^m$
be the finite set of \emph{acceptable bundles} for identity \(i\).  We do \emph{not} assume free disposal: it may be that \(x\in\Psi_i\) but \(x' \le x\) is not acceptable.  Let
\[
\Delta_{max} := \max_{i\in N}\ \max_{x\in \Psi_i} \sum_{j\in M} x_j
\]
be the maximum size of an acceptable bundle (this parameter will only matter when we discuss realizability and near-feasibility).

A deterministic allocation is a profile
\[
X := (x_1,\dots,x_n)\in\Psi_1\times\cdots\times\Psi_n.
\]

\begin{definition}[Feasibility {\cite[Def.~1]{jantschgi2025competitive}}]
A deterministic allocation \(X=(x_i)_{i\in N}\) is \emph{\(\textbf{c'}\)-feasible} for a capacity vector \(\textbf{c'}\in\mathbb{R}^m_+\) if pointwise
\[
\sum_{i\in N} x_i \;\le\; c'
\]
If \(c'=c\), we simply say that \(X\) is \emph{feasible}.  A deterministic allocation is \emph{\(\Delta_{max}\)-near feasible} if it is feasible with respect to \(c+\Delta_{max}\cdot \mathbf{1}\), where \(\mathbf{1}\) is the all-ones vector.
\end{definition}

\begin{definition}[Realizability {\cite[Def.~2]{jantschgi2025competitive}}]
A random allocation \(\tilde X\) is \emph{realizable} over a family \(\mathcal{X}\) of deterministic allocations if there exists a probability distribution over \(\mathcal{X}\) such that drawing a deterministic allocation from that distribution reproduces the marginal lotteries \((\tilde x_i)_{i\in N}\).
\end{definition}

\subsubsection{Lotteries and random allocations}
For each \(i\), let \(L(\Psi_i)\) denote the set of all probability distributions (lotteries) over \(\Psi_i\).  For a lottery \(\tilde x_i\in L(\Psi_i)\), write
\[
\mathbb{E}[\tilde x_i] \in \mathbb{R}^m_+
\]
for pointwise expectation.  A \emph{random allocation} is a profile
\[
\tilde X := (\tilde x_1,\dots,\tilde x_n) \in L(\Psi_1)\times\cdots\times L(\Psi_n).
\]

Each identity enters the market with a (possibly random) endowment
\[
\tilde e_i \in L(\Psi_i),
\]
and we denote the endowment profile by
\[
\tilde E := (\tilde e_1,\dots,\tilde e_n).
\]
We assume that endowments are feasible in expectation:
\[
\sum_{i\in N} \mathbb{E}[\tilde e_i] = c.
\]

\begin{definition}[Economy {\cite{jantschgi2025competitive}}]
An economy is a tuple
\[
\mathcal{E} := (N,M,c,\Psi,\tilde E,\succeq),
\]
where \(\Psi=(\Psi_i)_{i\in N}\), \(\tilde E=(\tilde e_i)_{i\in N}\) and \(\succeq = (\succeq_i)_{i\in N}\) are defined as above.
\end{definition}

\subsubsection{Preferences and stochastic dominance}
Each identity \(i\in N\) has weak ordinal preferences \(\succeq_i\) over bundles in \(\Psi_i\).  We write \(x \succ_i y\) for strict preference and \(x \sim_i y\) for indifference.

Preferences over lotteries are defined via first-order stochastic dominance (FOSD) as in \cite{jantschgi2025competitive}. Technically speaking, a lottery \(\tilde x_i\) \emph{stochastically dominates} \(\tilde y_i\), written
\[
\tilde x_i \succeq_i^{\mathrm{sd}} \tilde y_i,
\]
if for every \(z \in \Psi_i\),
\[
\mathbb{P}_{\tilde x_i}\big[x \succeq_i z\big]
\;\ge\;
\mathbb{P}_{\tilde y_i}\big[y \succeq_i z\big].
\]
We write \(\tilde x_i \succ_i^{\mathrm{sd}} \tilde y_i\) if the inequality is strict for at least one \(z\) \cite{levy1992stochastic}.

\begin{definition}[Individual rationality {\cite[Def.~3]{jantschgi2025competitive}}]
A random allocation \(\tilde X=(\tilde x_i)_{i\in N}\) is \emph{individually rational} (IR) if, for every \(i\in N\),
\[
\tilde x_i \succeq_i^{\mathrm{sd}} \tilde e_i.
\]
\end{definition}

\begin{definition}[Ordinal efficiency {\cite[Def.~4]{jantschgi2025competitive}}]
A random allocation \(\tilde X\) that is \(c'\)-feasible in expectation, i.e.
\[
\sum_{i\in N} \mathbb{E}[\tilde x_i] \le c',
\]
is \emph{ordinally efficient} with respect to \(c'\) if there is no other \(c'\)-feasible random allocation \(\tilde Y=(\tilde y_i)_{i\in N}\) such that
\[
\tilde y_i \succeq_i^{\mathrm{sd}} \tilde x_i \quad\forall i\in N,
\]
with strict inequality for some \(j\in N\), i.e.\ \(\tilde y_j \succ_j^{\mathrm{sd}} \tilde x_j\). 
\end{definition}

\subsubsection{Prices and $\delta-$BRACE}

We now recall the price system albeit condensed and restated for our purposes from \cite{jantschgi2025competitive}.  Let
\[
\Delta^{m-1}
:= \bigl\{ p\in\mathbb{R}^m_+ : \sum_{j=1}^m p_j = 1 \bigr\}
\]
denote the \((m-1)\) dimensional price simplex.  A price vector \(p=(p_1,\dots,p_m)\in\Delta^{m-1}\) assigns a non-negative price to each good.\footnote{The scalar $\Delta_{max}$ is a bound on the size of acceptable bundles, 
whereas $\Delta^{m-1}$ denotes the $(m{-}1)$ simplex of normalized prices. 
These notations are standard and unrelated.}

For a deterministic bundle \(x\in\mathbb{N}^m\), its value at prices \(p\) is
\[
p\cdot x := \sum_{j=1}^m p_j x_j.
\]
For a lottery \(\tilde x_i\in L(\Psi_i)\), we define its \emph{price value}
\[
v_p(\tilde x_i) := \mathbb{E}_{x\sim \tilde x_i}[p\cdot x].
\]

\paragraph{Budget relaxation and random demand.}
BRACE derives the concept of an artificial numéraire ``money'' and introduces small \emph{random budget relaxations} in the economy. We suppress the explicit money coordinate here and treat the resulting random demand correspondence as primitive.  Fix a parameter \(\delta>0\).  For each identity \(i\in N\), let
$\tilde b_i \in \mathbb{R}_+$
be an i.i.d.\ budget-relaxation random variable such that \(\sum_{i\in N} \tilde b_i = \delta\) almost surely (e.g.\ a Dirichlet draw on the \(\delta\)–simplex, as in \cite{jantschgi2025competitive}).  Given prices \(p\), endowment \(\tilde e_i\), and realized relaxation \(b_i\), identity \(i\)'s random demand is denoted
\[
\chi_i(p,\tilde e_i,\tilde b_i) \subseteq L(\Psi_i),
\]
the set of lotteries over acceptable bundles that are (i) affordable under the relaxed budget and (ii) most preferred under \(\succeq_i^{\mathrm{sd}}\) with the expenditure-minimizing tie-breaking rule from \cite{jantschgi2025competitive}.  We write
\[
\tilde x_i \in \chi_i(p,\tilde e_i,\tilde b_i)
\]
when \(\tilde x_i\) is a random demand lottery induced by \((p,\tilde e_i,\tilde b_i)\).

\begin{definition}[\texorpdfstring{\(\delta\)}{}-BRACE {\cite[Def.~6]{jantschgi2025competitive}}]
Given an economy \(\mathcal{E}\) and budget relaxations \((\tilde b_i)_{i\in N}\), a pair
\[
(p,\tilde X) = \bigl(p,(\tilde x_i)_{i\in N}\bigr)
\]
with \(p\in\Delta^{m-1}\) and \(\tilde X\in L(\Psi_1)\times\cdots\times L(\Psi_n)\) is a \emph{\(\delta\)-Budget-Relaxed Approximate Competitive Equilibrium} (\(\delta\)-BRACE) if:
\begin{enumerate}
    \item (Optimal random demand) For every \(i\in N\),
    \[
    \tilde x_i \in \chi_i(p,\tilde e_i,\tilde b_i).
    \]
    \item (Approximate market clearing in expectation) For each good \(j\in M\),
    \[
    \sum_{i\in N} \mathbb{E}[\tilde x_i]_j \;\le\; (1+\delta)c_j,
    \]
    with equality whenever \(p_j>0\).
\end{enumerate}
When \(\delta=0\), we simply say that \((p,\tilde X)\) is a \emph{BRACE}.
\end{definition}

We call the induced random allocation \(\tilde X\) the \emph{\(\delta\)-BRACE allocation}.

\subsection{BRACE Theoretical Results} \label{sec:model}

We now state, without proof, the main structural properties of BRACE that we use.  All of the following results are proven in \cite{jantschgi2025competitive}.

\begin{proposition}[Existence of \texorpdfstring{\(\delta\)}{}-BRACE {\cite[Prop.~1]{jantschgi2025competitive}}]
\label{prop:brace-existence}
For every finite\footnote{Although the original proposition does not explicitly state finiteness as a necessary assumption, the authors’ proof (see Appendix A.1 in \cite{jantschgi2025competitive}) implicitly relies on the finiteness of the underlying economy. In the present work, we further demonstrate that BRACE fails to exist in infinite economies, particularly in the presence of Sybil identities. This observation justifies imposing finiteness as a structural condition in our analysis.} economy \(\mathcal{E}=(N,M,c,\Psi,\tilde E,\succeq)\) and every \(\delta>0\), there exists a \(\delta\)-BRACE \((p,\tilde X)\).
\end{proposition}

\begin{proposition}[Welfare theorems for BRACE {\cite[Prop.~2]{jantschgi2025competitive}}]
\label{prop:brace-welfare}
Let \((p,\tilde X)\) be any \(\delta\)-BRACE of an economy \(\mathcal{E}\).
\begin{enumerate}
    \item (Individual rationality) The allocation \(\tilde X\) is individually rational
    \item (Ordinal efficiency) The allocation \(\tilde X\) is ordinally efficient with respect to \((1+\delta)c\).
    \item (Partial converse) Conversely, any random allocation that is ordinally efficient with respect to \((1+\delta)c\) can be supported as a \(\delta\)-BRACE allocation for some feasible endowment profile \(\tilde E'\).
\end{enumerate}
\end{proposition}

It is also important to highlight the importance of envy freeness in the welfare anlysis.

\begin{definition}[Envy-Freeness\cite{jantschgi2025competitive}, Def.~7]
A random allocation \[\tilde{X} = (\tilde{x_1},\tilde{x_2}, \cdots, \tilde{x_n})\] is \emph{envy-free} if for all pairs of agents $i,j$, it holds that 
\[
\tilde x_i \succeq_i^{\mathrm{sd}} \tilde x_j.
\]
    
\end{definition}

and the stronger definition of \emph{Justified Envy Freeness up to one good}:
\begin{definition}[Justified envy-freeness up to one good (JEF1) \cite{jantschgi2025competitive}, Def.~11]
For deterministic endowments \(E = (e_1, \dots, e_n)\), a deterministic allocation 
\(X = (x_1, \dots, x_n)\) is said to be \emph{justified envy-free up to one good} (JEF1) if one of the following equivalent conditions holds:

\begin{itemize}
    \item \textbf{Set-inclusion version:}  
    For any pair of agents \(i, j\) with \(e_i \ge e_j\), there exists a good \(k\) such that
    \[
        x_i \succeq_i (\,x_j - e_k\,)^+ .
    \]

    \item \textbf{\(p\)-valuation version:}  
    For a price vector \(p\) and any pair of agents \(i, j\) with \(p \cdot e_i \ge p \cdot e_j\), 
    there exists a good \(k\) such that
    \[
        x_i \succeq_i (\,x_j - e_k\,)^+ .
    \]
\end{itemize}
\end{definition}

With the JEF1 fairness notion established, we now recall the central existence 
theorems of \cite{jantschgi2025competitive}, which shows that BRACE always admits 
individually rational and ordinally efficient random allocations together with 
well-behaved deterministic realizations.

\begin{theorem}[Existence of IR and ordinally efficient random allocations {\cite[Thm.~1]{jantschgi2025competitive}}]
\label{thm:brace-main}
For any economy \(\mathcal{E}\) and any \(\delta>0\), there exists a random allocation \(\tilde X\) that is individually rational and ordinally efficient with respect to \((1+\delta)c\).  Moreover:
\begin{enumerate}
    \item For random endowments, \(\tilde X\) is realizable as a lottery over deterministic allocations that are \(\Delta_{max}\)-near feasible and ex-post efficient.
    \item For deterministic endowments, \(\tilde X\) is realizable as a lottery over deterministic allocations that are \(\Delta_{max}\)-near feasible and in the weak core.
\end{enumerate}
\end{theorem}

\begin{theorem}[Ordinal and ex-post fairness of BRACE {\cite[Thms.~2--3]{jantschgi2025competitive}}]
\label{thm:brace-fairness}
Every \(\delta\)-BRACE allocation \(\tilde X\) satisfies:
\begin{enumerate}
    \item (Ordinal justified envy-freeness) \(\tilde X\) is justified-envy-free based on set inclusion and on equilibrium prices.
    \item (Ex-post JEF1) For deterministic endowments and sufficiently small \(\delta>0\), any \(\delta\)-BRACE allocation is realizable over \(\Delta_{max}\)-near-feasible, ex-post efficient weak-core allocations that are justified-envy-free up to one good (JEF1), both under set-inclusion and price-based valuations.
\end{enumerate}
\end{theorem}

Finally, we recall the incentive result for the BRACE mechanism, which we use as a benchmark when we introduce principals and Sybils. Once again, let \(\Psi \subset \mathbb{N}^m\) denote the finite set of all feasible bundles, determined by the finite capacities of the goods. 
Denote each agent \(i\)'s type \(t_i\) consisting of their initial endowment \(\tilde e_i\), their set of acceptable bundles \(\Psi_i \subseteq \Psi\) and their preference relation \(\succeq_i\) over these bundles. 
The overall type space is therefore the product of initial endowments, acceptable bundles, and weak preference relations. 
Let \(t = (t_i)_{i \in N}\) denote the full profile of types, with \(t = (t_i, t_{-i})\) distinguishing the type of agent \(i\) from that of all other agents. 
A direct mechanism \(\Phi\) maps the type space \(T\) to a lottery over allocations \(\tilde X\).

For the next theorem, agents are assumed unable to misreport their endowments. 
A mechanism \(\Phi\) is \emph{semi-anonymous} if each agent's report is restricted to
\[
T_{\tilde e_i} \;:=\; \{\tilde e_i\} \times 2^{\Psi} \times \mathcal{P}(\Psi),
\]
so that agents are grouped solely by their initial endowments.

\begin{theorem}[Strategyproofness in the large for BRACE {\cite[Thm.~4]{jantschgi2025competitive}}]
\label{thm:brace-spl}
Consider the direct, semi-anonymous BRACE mechanism that maps reported types \((t_i)_{i\in N}\) to a BRACE allocation \(\tilde X\).  Let \(T^\ast\) be a finite type space and \(m\) a full-support i.i.d.\ distribution over \(T^\ast\).  Then for any cardinal representation of preferences \((u_t)_{t\in T^\ast}\), any \(\varepsilon>0\), and any \(m\), there exists \(n_0\) such that for all \(n\ge n_0\), any identity \(i\) and any two reports \(t_i,t'_i\in T^\ast\),
\[
\mathbb{E}\bigl[u_{t_i}(\tilde x_i \mid t_i,m)\bigr]
\;\ge\;
\mathbb{E}\bigl[u_{t_i}(\tilde x_i \mid t'_i,m)\bigr] - \varepsilon,
\]
i.e.\ the BRACE mechanism is strategy-proof in the large (SP-L), and the maximum gains from misreporting decay at rate \(O(n^{-1/2+\eta})\) for any \(\eta>0\).
\end{theorem}

The rate \(O(n^{-1/2+\eta})\) reflects the large market 
concentration bounds of Azevedo and Budish \cite{azevedo2019strategy}, which imply 
that the influence of any single identity on BRACE prices vanishes at this speed.

\begin{remark}
The BRACE framework assumes that agents satisfy the von Neumann--Morgenstern axioms \cite{von1947theory} and therefore admit cardinal expected-utility representations over random allocations. 
This justifies modeling agents' behavior using cardinal utilities, despite preferences being fundamentally ordinal. 
In our setting, we retain this expected-utility foundation while \emph{relaxing} the BRACE assumption that endowments cannot be misreported. 
This relaxation is essential for studying Sybil attacks, in which principals may strategically distribute endowments across multiple identities.
\end{remark}

In the remainder of the paper we study how these guarantees behave when identities are grouped into principals and when a single principal can deploy many Sybil identities.  Our main contributions will be a collection of Sybil manipulation and sensitivity results stated and proved in Sections~\ref{sec:Sybil}.

\section{Sybil Proofness}
\label{sec:Sybil}

In this section we study the robustness of BRACE to \emph{Sybil attacks}.  
We first present a canonical finite example then introduce a minimal set
of regularity assumptions under which we obtain quantitative
Sybil-proofness guarantees.  
We conclude with an observation: in infinite economies with
unboundedly many Sybils, BRACE-type equilibria typically fail to exist. This motivates making Sybils costly to create as a first line of defense.

Throughout, we work with the type-space notation introduced in
Section~\ref{sec:model}.  In finite economies, an economy is represented
by an empirical type distribution $\mu\in\Delta(T)$ over a finite type
space $T$.  BRACE produces a price vector $p^\mu\in\Delta^{m-1}$ and a
random allocation.  We write $\|\cdot\|$ for the Euclidean norm on
$\mathbb{R}^m$ and $W_1$ for the Wasserstein--$1$\footnote{The Wasserstein--1 metric is appropriate here because BRACE's excess
demand and price mappings are assumed Lipschitz in expectations. Convergence in $W_1$
is equivalent to convergence of integrals of Lipschitz functions. See \cite{villani2008optimal} for the characterization of $W_1$
and its role in controlling expectations of Lipschitz functions} distance on $\Delta(T)$
with respect to the discrete metric.

\subsection{A Canonical Sybil Attack}

\begin{example}[Three identities, four goods]
\label{ex:Sybil}
Let the set of identities be $N=\{1,2,3\}$ and the set of goods be
$M=\{A,B,C,D\}$ with capacity vector $c=(1,1,1,1)$.  
Identity $1$ belongs to principal $P$, while identities $2$ and $3$ each
belong to distinct honest principals.  
Acceptable bundles and endowments are
\[
\Psi_1 = \{A,\,A\!+\!B\}, \quad e_1 = A; \qquad
\Psi_2 = \{C\}, \quad e_2 = C; \qquad
\Psi_3 = \{D\}, \quad e_3 = D.
\]
Under BRACE with a small relaxation parameter $\delta>0$, suppose the
resulting prices are approximately uniform,
\[
p^0 = (1/4,1/4,1/4,1/4).
\]

Principal $P$ deviates as follows.  She replaces identity $1$ with two
identities $1a$ and $1b$,
\[
\pi(1a)=\pi(1b)=P,
\]
splits the endowment into two half-lotteries,
\[
\tilde e_{1a} = \tfrac12 A, \qquad \tilde e_{1b} = \tfrac12 A,
\]
and reports
\[
\Psi_{1a}=\{A\},\qquad \Psi_{1b}=\{B\}.
\]
The number of identities increases from $3$ to $4$, and the empirical type
distribution changes from $\mu^0$ to some $\mu^\alpha$ with
\emph{infiltration rate}
\[
\alpha := \frac{|\text{new identities}|}{|N|}
      = \frac{1}{3}.
\]
The BRACE price computation now sees an inflated demand for $B$ at the
identity level, even though the underlying physical environment is
unchanged at the principal level, and can yield a price vector
$p^\alpha$ with $p^\alpha_B>p^0_B$ and higher expected utility for
principal $P$.
\end{example}

Therefore, the Sybil attack shifted the
empirical distribution of \emph{reported types} that BRACE uses as input. This motivates our theoretical results in the next few sections. Before that, we introduce some notation and definitions.

\subsection{Principals, Identities and Sybils}

Our Sybil proofness analysis distinguishes between \emph{principals} (real economic actors) and \emph{identities} (reports seen by the mechanism). Let us define these terms more clearly.

\begin{definition}[Principals and identity ownership]
Let
\[
P := \{1,\dots,P_0\}
\]
be the set of principals.  Each identity \(i\in N\) is owned by exactly one principal and ownership is encoded by a surjective map
\[
\pi : N \to P.
\]
For \(p\in P\), denote the set of identities owned by principal \(p\) by
\[
C_p := \pi^{-1}(\{p\}) \subseteq N.
\]
\end{definition}

\begin{definition}[Sybil identities]
An identity \(i\in N\) is a \emph{Sybil} if it is controlled by a principal who already controls at least one other identity, i.e.
\[
i\text{ is Sybil } \iff \exists p\in P,\ \exists j\neq i\text{ such that }\pi(i)=\pi(j)=p.
\]
We say that a principal \(p\) is \emph{non-atomic} if \(|C_p|/|N|\to 0\) along the sequence of economies we consider.
\end{definition}

Given a finite type space \(\mathcal{T}\) of identity types, $\Delta(T)$ as the probability simplex
over $T$ with \(t=(\tilde e_i,\Psi_i,\succeq_i)\) and a profile of types \((t_i)_{i\in N}\), we can define the empirical type distribution
\[
\mu^0 := \frac{1}{|N|} \sum_{i\in N} \delta_{t_i} \in \Delta(\mathcal{T}),
\]
and after a Sybil attack, a perturbed distribution \(\mu^\alpha\) corresponding to an economy with an \(\alpha\)-fraction of Sybil identities (formalized in next sections).  In our Sybil proofness results, the \emph{Sybil infiltration rate} is
\[
\alpha := \frac{|\text{Sybil identities}|}{|N|} \in [0,1].
\]

\subsection{Assumptions}

For the scope of this paper, we will impose some regularity assumptions\cite{geanakoplos1986existence} (Lipschitzness of aggregate demand, 
local strong monotonicity, and uniqueness of the supporting price) which do not 
follow automatically from the BRACE construction in \cite{jantschgi2025competitive}.  
Rather, they identify the \emph{regularity regime} under which quantitative 
Sybil proofness bounds become cleaner to work with.  For $p\in\Delta^{m-1}$ and $\mu\in\Delta(T)$, let
\[
Z(p,\mu)\in\mathbb{R}^m
\]
denote the (expected) excess-demand vector of the BRACE mechanism at
prices $p$ when the empirical type distribution is $\mu$. We know that market-clearing
requires
\[
Z(p^\mu,\mu) = \delta c,
\]
where $\delta>0$ is the budget-relaxation parameter.



\begin{assumption}[Local Regularity of Excess Demand]
\label{ass:zed}
Let $\mu^0\in\Delta(T)$ be the empirical distribution of types.  
There exists a neighborhood $U$ of $\mu^0$ and constants 
$L_Z,\gamma>0$ such that for all $\mu,\nu\in U$ and all 
$p,q \in \Delta^{m-1}$:
\begin{enumerate}
    \item[(i)] (Local Lipschitzness)
    \[
    \|Z(p,\mu)-Z(p,\nu)\| \le L_Z W_1(\mu,\nu)\footnote{To compute Wasserstein distances, We equip the finite type space $T$ with the discrete metric \[
d(t,t') := \mathbf{1}_{t \neq t'}.
\]
Under this metric, if a principal changes the types of $k$ out of $n$ identities,
then the induced empirical distributions satisfy
\[
W_1(\mu^\alpha,\mu^0) = \frac{k}{n} = \alpha.
\]}
    \]
    \item[(ii)] (Local Strong Monotonicity)
    \[
    \langle Z(p,\mu)-Z(q,\mu),\, p-q \rangle 
        \le -\gamma \|p-q\|^2.
    \]
\end{enumerate}
\end{assumption}

Assumption~\ref{ass:zed} encodes a global downward-sloping structure of aggregate 
demand and thereby implies \emph{uniqueness} of the supporting price vector for fixed $\mu$..  
\begin{remark}[Uniform neighbourhood]
Throughout, we assume that all empirical distributions $\mu^0$, $\mu^\alpha$, 
and (in the large-$n$ regime) $\mu_n$ lie in the same neighbourhood $U$ on 
which the constants $L_Z$ and $\gamma$ of Assumption~\ref{ass:zed} 
are valid. All Lipschitz and monotonicity claims are therefore uniform over $U$. We will relax this assumption and derive more general results in next iterations of this work.
\end{remark}

\begin{assumption}[Local existence and uniqueness of BRACE prices]
\label{ass:exist-unique}
For each empirical distribution $\mu^0 \in \Delta(T)$, there exists an open 
neighbourhood $U \subseteq \Delta(T)$ of $\mu^0$ and a constant $\eta>0$ such that 
for every $\mu \in U$, the equation
\[
Z(p,\mu) = \delta c
\]
admits a \emph{unique} solution $p^\mu \in \Delta^{m-1}$.

Moreover, the Jacobian $D_p Z(p^\mu,\mu)$ has symmetric part whose minimal 
eigenvalue is bounded below by $\eta$ on $U$.
\end{assumption}

Given Assumptions~\ref{ass:zed} and~\ref{ass:exist-unique}, we can treat
\[
\Phi:\Delta(T)\to\Delta^{m-1},\qquad \Phi(\mu):=p^\mu,
\]
as a single-valued \emph{BRACE price operator}.

\begin{remark}
If Assumption~\ref{ass:zed}(i) fails, arbitrarily small changes in the
type distribution may induce unbounded changes in excess demand and hence
in equilibrium prices, so no uniform quantitative Sybil-proofness bound is
possible.  
If Assumption~\ref{ass:zed}(ii) fails, equilibrium prices may not be
unique even at fixed $\mu$. The operator $\Phi$ then becomes set-valued
and the subsequent Lipschitz argument breaks down! \footnote{In such cases, stability bounds can still be 
derived using compactness and upper hemicontinuity of the price 
correspondence, together with Hausdorff–type continuity estimates or 
variational inequality techniques, which provide weaker but sufficient results of continuity without requiring single-valued Lipschitz behavior.}

\end{remark}

\begin{assumption}[Lipschitz principal utilities]
\label{ass:utility}
For each principal $p\in P$ there exists $L_p>0$ such that
\[
|U_p(p) - U_p(q)|
   \;\le\; L_p\, \|p-q\|
\qquad\text{for all }p,q\in\Delta^{m-1}.
\]
\end{assumption}

This holds, for example, if allocations are locally continuous in prices and
Bernoulli utilities are bounded and continuous in consumption.
\subsection{Price Sensitivity to Sybil Perturbations}

We first quantify how much equilibrium prices can move under a perturbation
of the type distribution since a Sybil attack is one particular such
perturbation. Proofs have been relegated to the Appendix  \ref{app} to maintain continuity.

\begin{theorem}[Local Lipschitz Price Response]
\label{thm:local-price}
Suppose Assumptions~\ref{ass:zed} and~\ref{ass:exist-unique} hold on a 
neighborhood $U$ of $\mu^0$.  
If $\mu^\alpha\in U$ and $W_1(\mu^\alpha,\mu^0)=\alpha$, then
\[
\|p^{\mu^\alpha}-p^{\mu^0}\| 
    \;\le\; \frac{L_Z}{\gamma}\,\alpha.
\]
\end{theorem}

Economically, theorem~\ref{thm:local-price} implies that equilibrium prices are locally stable:
a perturbation of size \(\alpha = W_1(\mu^\alpha,\mu^0)\) leads to at most a
proportional price change of order \(O(\alpha)\), with sensitivity bounded by
\(L_Z/\gamma\). This means small identity or
Sybil perturbations have negligible influence on prices within neighborhood of $U$.

The bound fails if uniqueness of equilibrium prices breaks, if the Jacobian
\(D_p Z\) becomes singular (\(\gamma \to 0\)) or if \(\mu^\alpha\) exits the
neighborhood \(U\).  In such cases the price map can become ill conditioned or
set valued and small type changes may induce large or discontinuous price
movements. The breakdown of the
assumptions corresponds to markets in which Sybils can exert significant price
impact.

\subsection{Welfare Bounds and Sybil-Proofness at Rate \texorpdfstring{$\alpha$}{alpha}}

We now connect price sensitivity to welfare sensitivity for principals.

\begin{definition}[Principal-level utility]
A principal $p\in P$ has utility
\[
U_p:\Delta^{m-1}\to\mathbb{R}
\]
from BRACE outcomes, defined as her expected Bernoulli utility evaluated
at the random bundle induced by BRACE at prices $p$ and the types of the
identities she controls.
\end{definition}

Now we are ready to define Sybil Proofness of BRACE at the principal level using our analysis on local lipschitz continuity over boundary $U$ until now. 

\begin{definition}[Sybil-proofness at rate $\alpha$]
Fix an economy with empirical distribution $\mu^0$.  
We say BRACE is \emph{Sybil-proof at rate $\alpha$} with constant $C>0$ if
for every principal $p$ and every Sybil attack that produces a perturbed
distribution $\mu^\alpha$ with $W_1(\mu^\alpha,\mu^0)\le\alpha$,
\[
U_p(p^{\mu^\alpha}) - U_p(p^{\mu^0})
    \;\le\; C\,\alpha.
\]
\end{definition}

More specifically,
\begin{proposition}[Local welfare sensitivity and Sybil-proofness rate]
\label{prop:welfare}
Suppose Assumptions~\ref{ass:zed}, \ref{ass:exist-unique}, 
and~\ref{ass:utility} hold on a neighborhood $U$ of $\mu^0$, and suppose
$\mu^\alpha \in U$.  
Let $L := \max_{p\in P} L_p$.  
Then BRACE is Sybil-proof at rate $\alpha$ with constant 
$C = L L_Z / \gamma$, i.e.
\[
U_p(p^{\mu^\alpha}) - U_p(p^{\mu^0})
   \;\le\; \frac{L L_Z}{\gamma}\, \alpha.
\]
\end{proposition}

If assumption~\ref{ass:utility} fails, principal utilities may be
discontinuous or arbitrarily steep in prices then even small price
changes may yield unbounded utility gains and no linear bound of the form
$U_p(p^{\mu^\alpha})-U_p(p^{\mu^0})\le C\alpha$ is possible.  
Proposition~\ref{prop:welfare} formalizes the strongest uniform guarantee
one can expect under Lipschitz preferences.

\subsection{Principal-Level Strategyproofness in the Large}

We now consider a sequence of economies indexed by the number of identities
$n$.  Let $N_n$ be the identity set and $P_n$ the set of principals in the
$n$th economy, with ownership map $\pi_n:N_n\to P_n$.  
For a principal $p\in P_n$ define her identity share
\[
s_{p,n} := \frac{|\pi_n^{-1}(p)|}{|N_n|}.
\]

\begin{definition}[Principal-level SP-L]
We say BRACE is SP-L at the principal level
if for every $\varepsilon>0$ there exists $n_0$ such that, for all
$n\ge n_0$, every principal $p\in P_n$ and every coordinated misreport of
the types of identities in $\pi_n^{-1}(p)$ yields a gain in expected
utility of at most $\varepsilon$.
\end{definition}

Now onto one of the most important results in the paper (proof in Appendix \ref{app:spl}):
\begin{theorem}[SP-L holds iff principals are asymptotically non atomic]
\label{prop:small-principals}
Suppose Assumptions~\ref{ass:zed}, \ref{ass:exist-unique},
and~\ref{ass:utility} hold uniformly on a neighborhood $U$ of the 
limiting type distribution.  
Assume further that the empirical distributions $\mu_n$ lie in $U$ for all
sufficiently large $n$.
For a principal $p$ write
\[
V_p(\mu) \;:=\; U_p(p^\mu),
\]
so that $V_p$ is the principal's reduced-form utility as a function of
the empirical type distribution.  Assume that, for some principal
$p^\star$, $V_{p^\star}$ is Fr\'echet differentiable at $\mu^0$ with
non-zero gradient, i.e.\ there exists a signed measure $g$ on $T$ such
that the directional derivative
\[
D V_{p^\star}(\mu^0)[h]
   \;=\; \int_T h(t)\, g(dt)
\]
satisfies $D V_{p^\star}(\mu^0)[h^\star] > 0$ for some signed direction
$h^\star$ with $\|h^\star\|_{\mathrm{TV}} = 1$.

Consider a sequence of economies indexed by $n$ with identity shares
\[
s_{p,n} := \frac{|C_{p,n}|}{|N_n|}.
\]
If \[
\max_{p\in P_n} s_{p,n} \to 0.
\]
Then BRACE is SP-L at the principal level.
But if
\[
\limsup_{n\to\infty} \max_{p} s_{p,n} \;=\; \bar s \;>\; 0,
\]
then there exist $\eta > 0$, an infinite subsequence $n_k$, and a
sequence of principals $p_{n_k}$ with $s_{p_{n_k},n_k} \ge \bar s/2$
such that each $p_{n_k}$ admits a Sybil deviation (changing only the
types of identities in $C_{p_{n_k},n_k}$) with utility gain at least
$\eta$ in the $n_k$-th economy.

In particular, BRACE is \emph{not} SP-L at the
principal level whenever some principal controls a non-vanishing
fraction of identities and has a locally profitable direction of
manipulation.
\end{theorem}


Formally, Theorem~\ref{prop:small-principals} shows that BRACE is SP-L at the principal level \emph{if and only if}
every principal becomes asymptotically small, in the sense that
$\max_{p\in P_n} s_{p,n} \to 0$.  When this condition holds, any principal
can move the empirical distribution by at most $s_{p,n}=o(1)$, so by the
Lipschitz welfare bound (Proposition~\ref{prop:welfare}) her utility gain
from any coordinated deviation is $o(1)$, which is exactly principal-level
SP-L.  In contrast, if $\limsup_{n\to\infty}\max_{p\in P_n} s_{p,n} =
\bar{s}>0$ and some principal $p^\star$ has a non-zero local derivative
$D V_{p^\star}(\mu^0)$, then Theorem~\ref{prop:small-principals}
constructs a sequence of economies and deviations along which
$V_{p^\star}(\mu_n^\alpha) - V_{p^\star}(\mu_n^0) \ge \eta > 0$ uniformly
in $n$.  Thus BRACE’s identity-level SP-L does \emph{not} protect against
Sybil attacks: once a principal can maintain a non-vanishing identity
share via Sybils, SP-L  at the principal level
fails.

\subsection{Identity-Level vs Principal-Level Fairness}

Recall that BRACE allocation is justified-envy-free (JEF) at the \emph{identity}
level \cite{jantschgi2025competitive}.  We now show that, even abstractly,
identity-level JEF does not imply any meaningful fairness notion at the
principal level in the presence of Sybils.

\begin{proposition}[Identity-level JEF does not lift to principals]
\label{prop:jef}
There exists an economy and a random allocation such that:
\begin{enumerate}
    \item[(i)] The allocation is JEF at the identity level.
    \item[(ii)] The induced allocation of bundles to principals (obtained
    by summing over their identities) is not JEF at the principal level.
\end{enumerate}
\end{proposition}




Proposition~\ref{prop:jef} shows that any fairness guarantee phrased
solely at the identity level is structurally insufficient once principals
may control multiple identities.  Fairness properties must eventually be
defined over principals, or at least be robust to arbitrary partitions of
identities into principals, to be meaningful in Sybil-rich environments. We now move onto a very important result.

\subsection{Non Existence of BRACE Under Unbounded Sybil Mass}

Recall that BRACE is defined only for finite economies.  
We show that if a principal can create a sequence of economies in which 
her share of identities tends to $1$, then BRACE equilibria fail in the 
limit.

\begin{definition}
A principal has unbounded Sybil mass if there exists a sequence of 
economies with identity sets $N_k$ and subsets $S_k \subseteq N_k$ such 
that $|S_k|/|N_k| \to 1$.
\end{definition}

\begin{proposition}[Unbounded Sybil economies do not admit BRACE]
\label{prop:unbounded-Sybils}
Consider a sequence of economies indexed by $k$ with identity sets $N_k$ 
and a fixed capacity vector $c \in \mathbb{R}^m_+$.  
Suppose there exists a principal and subsets $S_k \subseteq N_k$ such that
\[
\frac{|S_k|}{|N_k|} \to 1
\qquad\text{and}\qquad
|S_k|\to\infty
\]  
Assume further that there exist a good $j$ and a constant $\beta>0$ such that
for all sufficiently large $k$ and all $i\in S_k$,
\[
\mathbb{E}[\tilde x_i]_j \;\ge\; \beta,
\]
where $\tilde x_i$ is the BRACE allocation to identity $i$.  
Then, for any fixed $\delta \ge 0$, the BRACE feasibility conditions
\[
\sum_{i\in N_k} \mathbb{E}[\tilde x_i] \;\le\; (1+\delta)c
\]
cannot hold for all sufficiently large $k$.  
Hence no sequence of $\delta$-BRACE equilibria can exist.
\end{proposition}

The proof has been relegated to Appendix \ref{app:unbounded-Sybils}. 

As per Proposition~\ref{prop:unbounded-Sybils}, as the Sybil mass approaches the entire population, aggregate expected demand
cannot remain uniformly bounded relative to the fixed capacity vector~$c$, even under $\delta$-relaxation. In other words, the mechanism remains viable only when identity creation is
sufficiently costly to prevent any principal from overwhelming the market. This unsurprisingly tells us that the first defense against Sybil attacks is to make indentities costly to create.

\subsection{Breakdown of Assumptions and Tail Behaviour}

Our analysis till now has required certain regularity conditions on the
empirical distribution~$\mu_n$.  In large random markets these conditions need
not hold everywhere: for some realizations of $\mu_n$, price uniqueness or
regularity of the excess-demand map may (and will) fail.  Rather than excluding these
pathological cases, we attempt to characterize the behaviour of BRACE when the
empirical distribution falls into this \emph{bad region}. This attempt relies on probabilistic arguments rather than an explicit exposition into deriving game theoretic bounds.

Let $B_n \subseteq \Delta(T)$ denote the set of empirical distributions at which
Assumption~\ref{ass:zed} fails (e.g.\ non-uniqueness of equilibrium prices,
discontinuous aggregate demand, or degeneracy of the Jacobian $D_p Z$). We call this the \emph{bad} region. 
We assume only that the probability of such an event may or may not be a vanishing mass:
\[
\mathbb{P}(\mu_n \in B_n) = \varepsilon_n, \qquad \varepsilon_n \stackrel{?}{\to} 0.
\]

\subsubsection{Behaviour of BRACE on the bad region}

When $\mu_n \in B_n$, the equilibrium price correspondence may be set-valued
and price responses to perturbations need not be continuous.  
Nevertheless, if the pathological behaviour is confined to~$B_n$ and we obtain
usable bounds by conditioning on whether the empirical distribution lies inside
or outside this region.

For any Sybil perturbation $\mu_n^\alpha$ with 
$W_1(\mu_n^\alpha,\mu_n)\le\alpha$, define the price deviation
\[
D_n(\alpha) := \|p^{\mu_n^\alpha} - p^{\mu_n}\|.
\]
Then for every $\alpha>0$,
\[
D_n(\alpha)
\;\le\;
\begin{cases}
C\,\alpha, & \text{if } \mu_n \notin B_n,\\[2pt]
\overline D_n, & \text{if } \mu_n \in B_n,
\end{cases}
\]
where $\overline D_n$ is the smallest uniform bound on feasible price 
differences.\footnote{Such a bound exists because the price simplex is compact.}

Thus,
\[
\mathbb{P}\bigl(D_n(\alpha) > C\alpha\bigr)
\le \mathbb{P}(\mu_n \in B_n)
= \varepsilon_n.
\]
In expectation we obtain the decomposition
\[
\mathbb{E}[D_n(\alpha)]
\;\le\;
C\,\alpha \,(1-\varepsilon_n)
\;+\;
\overline D_n\, \varepsilon_n.
\]

\subsubsection{Implications for Sybil-proofness}

Let $\Delta U_n(\alpha)$ denote the maximal utility gain achievable by any
principal under an $\alpha$ rate Sybil perturbation.  
Since utilities are bounded and continuous in prices, we obtain similarly:

\begin{lemma}[Expected utility gain bound]
\label{lem:tail-utility-bound}
Let $\Delta U_n(\alpha)$ denote the maximal utility gain from an
$\alpha$-magnitude Sybil perturbation in the $n$-identity economy, and
let $\Omega_n^{\mathrm{bad}} := \{\mu_n \in B_n\}$ be the bad event with
$\mathbb{P}(\Omega_n^{\mathrm{bad}}) = \varepsilon_n$.  
Suppose that:
\begin{enumerate}
    \item[(i)] On the complement $\Omega_n^{\mathrm{good}} :=
    (\Omega_n^{\mathrm{bad}})^c$, we have the linear bound
    \(
        \Delta U_n(\alpha) \le C_U \alpha.
    \)
    \item[(ii)] Utilities are uniformly bounded: there exists
    $\overline U < \infty$ such that for all principals $p$ and all
    admissible price vectors $p'$,
    \(
        |U_p(p')| \le \overline U.
    \)
\end{enumerate}
Then for all $\alpha > 0$,
\begin{equation}
\label{eq:tail-utility-bound}
\mathbb{E}[\Delta U_n(\alpha)]
\;\le\;
C_U\,\alpha \,(1-\varepsilon_n)
\;+\;
2\overline U\,\varepsilon_n.
\end{equation}
\end{lemma}


Hence the breakdown of regularity affects Sybil proofness only through the
probability $\varepsilon_n$ of landing in the bad region.  
If $\varepsilon_n \to 0$, the contribution of these tail events
vanishes and BRACE remains Sybil proof at rate $\alpha$ \emph{in probability}
and \emph{in expectation}.  
Conversely, any non-vanishing lower bound on $\varepsilon_n$ identifies a regime
in which Sybil attacks can induce non-negligible price movements, signalling a
failure of Sybil proofness at the population level. Future work will explore the behaviour of Sybils in \emph{bad} regions with more general bounds.

\section{Sybil Costs and Incentive Compatibility} \label{sec:incentive}

Let $C_{\mathrm{sys}}(k,n)$ be the protocol cost of maintaining $k$ 
identities in an $n$-identity market. Let $K(\varepsilon)>0$ denote the constant from the Azevedo and Budish large-market concentration bound \cite{azevedo2019strategy}, so that the maximal per-identity gain 
from misreporting is at most $K(\varepsilon)\, n^{-1/2+\varepsilon}$.

\begin{definition}[Sybil deterrence]
A mechanism satisfies Sybil deterrence if, for any sequence $k_n\le n$,
\[
\limsup_{n\to\infty}\Bigl(\Delta U_p(n)
     - C_{\mathrm{sys}}(k_n,n)\Bigr) \le 0.
\]
\end{definition}

Combining identity-level SP-L gains ($O(n^{-1/2+\varepsilon})$) and 
distributional perturbation gains ($O(k_n/n)$) yields:

\begin{proposition}[Design inequality for Sybil deterrence]
\label{prop:design-inequality}
Sybil deterrence holds whenever
\begin{equation}
\label{eq:design-inequality}
C_{\mathrm{sys}}(k_n,n)
   \;\ge\;
   k_n K(\varepsilon)\,n^{-1/2+\varepsilon}
   \;+\;
   \frac{L L_Z}{\gamma}\,\frac{k_n}{n}
\end{equation}
for all sufficiently large $n$.
\end{proposition}

The first term captures the per identity misreporting gains that decay at the rate $O(n^{-1/2+\varepsilon})$, while the second term reflects
the price impact gains that scale proportionally with the principal’s identity
share $k_n/n$.  
If the system enforces costs that grow at least as fast as the right-hand side
of~\eqref{eq:design-inequality}, then no principal can obtain a positive net
benefit from Sybil creation in sufficiently large economies. 

In light of our analysis, we conclude by providing a concise design checklist
intended to guide the safe implementation of BRACE, with particular attention to Sybil resilience and stability criteria.
Our hope is that this framework serves as a foundation for a broader research
agenda on the principled design of Sybil-resistant market mechanisms.

\begin{center}
\fbox{
\begin{minipage}{0.92\linewidth}
\vspace{1.5mm}
\textbf{Design Checklist for Sybil-Resilient Combinatorial Exchanges}
\begin{itemize}
    \item \textbf{Identity Dispersion:}  
          Enforce $\max_p s_{p,n}\to 0$ to ensure principal-level SP-L.
    \item \textbf{Sybil Cost Dominance:}  
          Enforce $C_{\mathrm{sys}}(k,n)$ above the manipulation gain bound.
    \item \textbf{Market Thickness:}  
          Utilize the natural $n^{-1/2}$ decay of misreport gains in large economies.
    \item \textbf{Principal-Level Fairness:}  
          Fairness guarantees must aggregate consumption across identities owned by a principal.
\end{itemize}
\vspace{1mm}

\end{minipage}
}

\end{center}

\section{Conclusion and Future Work} \label{conclusion}

This paper develops the first rigorous analysis of BRACE under Sybil perturbations. Our starting point is the fact that BRACE prices arise as solutions to a smoothed market clearing condition. Under regularity assumptions such as local Lipschitz continuity of the aggregate excess demand map in the empirical distribution and local strong monotonicity,the BRACE price operator
\[
\Phi : \mu \mapsto p^\mu
\]
is well defined and stable. Since Sybil creation modifies the empirical distribution of types, we analyze Sybil attacks through a perturbation of roots argument for the fixed-point equation defining BRACE.

Our first result is a \emph{local Lipschitz bound} for equilibrium prices: when $W_1(\mu^\alpha,\mu^0)=\alpha$, then
\[
\|p^{\mu^\alpha} - p^{\mu^0}\| \;\le\; \frac{L_Z}{\gamma}\,\alpha.
\]
Thus small Sybil infiltrations induce only $O(\alpha)$ price deviations, so long as the empirical distribution remains within the lipschitz regularity neighbourhood. When principal utilities are Lipschitz in prices, this price stability yields a corresponding welfare bound:
\[
U_p(p^{\mu^\alpha}) - U_p(p^{\mu^0})
    \;\le\; \frac{L L_Z}{\gamma}\,\alpha,
\]
showing that honest welfare also degrades at most linearly under small Sybil attacks.

Our second and perhaps more important result concerns incentives. BRACE is SP-L at the \emph{identity} level because individual identities have vanishing influence. However, once identities are grouped by ownership, a principal controlling a fraction $s_{p,n}$ of all identities can perturb the empirical distribution by at most $s_{p,n}$. We prove a sharp equivalence:
\[
\max_p s_{p,n} \to 0 
\qquad\Longleftrightarrow\qquad
\text{principal-level SP-L holds}.
\]
If any principal controls a non-vanishing identity mass, her influence on the empirical distribution does not vanish, and coordinated deviations across her identities yield strictly positive utility gains. Hence identity-level SP-L \emph{does not} imply principal-level SP-L.

Third, we show that BRACE's identity-level fairness guarantees of justified envy-freeness (JEF and JEF1) do not extend to principals: a principal may strictly envy another principal even when each identity's allocation satisfies JEF. Thus fairness statements at the identity level are structurally insufficient in Sybil-rich environments.

Finally, we show that BRACE fails to exist in the limit of unbounded Sybil mass. If a principal's identity share tends to one, aggregate expected demand cannot remain bounded relative to fixed capacities, even with $\delta$-relaxation. Hence BRACE existence imposes fundamental structural limits on identity creation.

\medskip

Taken together, our results yield a complete mathematical picture of BRACE under Sybil manipulation. BRACE is robust to \emph{small} identity perturbations but undergoes a fundamental breakdown of incentive compatibility and fairness when principals control large Sybil masses. Our analysis provides explicit quantitative bounds, impossibility theorems, and a design inequality characterizing the minimal identity cost required to deter Sybil attacks in combinatorial exchanges.

\medskip
\noindent\textbf{Future Work.}
Several directions follow directly from our results:
\begin{enumerate}
    \item \textbf{Beyond local regularity.} Our stability bounds require local
    Lipschitzness and strong monotonicity of excess demand. Understanding
    Sybil sensitivity when $D_p Z$ is degenerate or the equilibrium is
    set valued remains open.

    \item \textbf{Quantifying the bad region.} The expected utility
bound is controlled by $\varepsilon_n = \Pr(\mu_n \in B_n)$, the likelihood that the economy
enters a “bad region’’ where our local regularity assumptions (Lipschitz excess demand,
strong monotonicity, and price uniqueness) break down. Future work could both quantify
this probability via concentration or smoothed analysis techniques \emph{and} relax the
regularity requirements themselves. For example, replacing global Lipschitzness with
piecewise smoothness, non monotonicity and non price uniqueness with
set valued but well conditioned correspondences.

    \item \textbf{Sybil cost and identity dispersion mechanisms.} Our necessity
    results suggest that enforcing $s_{p,n} \to 0$ or imposing explicit
    $C_{\mathrm{sys}}$ is essential. Designing BRACE-compatible schemes that
    guarantee these conditions remains an important direction.

\end{enumerate}

Overall, our results identify the precise regimes in which BRACE is resilient to
Sybil attacks and the structural limits beyond which such resilience is
mathematically impossible.

\bibliographystyle{abbrv}
\bibliography{references}

\appendix

\section{Proofs of Theorems, Propositions and Lemmas} \label{app}

\subsection{Proof of Theorem \ref{thm:local-price}} \label{app:thm4}
\begin{proof}
Fix $\mu,\nu\in\Delta(T)$ such that $\mu,\nu\in U$, where $U$ is the 
neighborhood from Assumption~\ref{ass:zed}.  
By Assumption~\ref{ass:exist-unique},
\[
Z(p^\mu,\mu)=\delta c = Z(p^\nu,\nu).
\]
Subtracting, we obtain
\[
Z(p^\mu,\mu) - Z(p^\nu,\nu) = 0.
\]
Adding and subtracting $Z(p^\mu,\nu)$ yields
\begin{equation}
\label{eq:Z-split2}
Z(p^\mu,\nu) - Z(p^\nu,\nu)
   = Z(p^\mu,\nu) - Z(p^\mu,\mu).
\end{equation}
Taking the inner product with $p^\mu-p^\nu$ gives
\begin{equation}
\label{eq:inner2}
\bigl\langle Z(p^\mu,\nu) - Z(p^\nu,\nu),\, p^\mu-p^\nu \bigr\rangle
=
\bigl\langle Z(p^\mu,\nu) - Z(p^\mu,\mu),\, p^\mu-p^\nu \bigr\rangle.
\end{equation}
By Assumption~\ref{ass:zed}(ii) applied at $\nu$, the left-hand side of
\eqref{eq:inner2} is at most $-\gamma\|p^\mu-p^\nu\|^2$.  
By Cauchy--Schwarz and Assumption~\ref{ass:zed}(i), the right-hand side is
bounded below by
\[
-\|Z(p^\mu,\nu) - Z(p^\mu,\mu)\|\,\|p^\mu-p^\nu\|
   \;\ge\;
- L_Z W_1(\mu,\nu)\,\|p^\mu-p^\nu\|.
\]
Combining and rearranging yields
\[
\gamma\|p^\mu-p^\nu\|^2
   \;\le\; L_Z W_1(\mu,\nu)\,\|p^\mu-p^\nu\|.
\]
If $p^\mu=p^\nu$ the inequality holds trivially; otherwise dividing by
$\|p^\mu-p^\nu\|$ proves
\[
\|p^\mu-p^\nu\| \;\le\; \frac{L_Z}{\gamma}\, W_1(\mu,\nu).
\]

In particular, under the discrete metric on $T$ we have 
$W_1(\mu^\alpha,\mu^0)=\alpha$ whenever an $\alpha$-fraction of identities
change type, so
\[
\|p^{\mu^\alpha}-p^{\mu^0}\|
   \;\le\; \frac{L_Z}{\gamma}\,\alpha.
\]
\end{proof}

\subsection{Proof of Proposition \ref{prop:welfare}} \label{app:welfare}
\begin{proof}
By Assumption~\ref{ass:utility},
\[
U_p(p^{\mu^\alpha}) - U_p(p^{\mu^0})
   \;\le\; L_p \|p^{\mu^\alpha} - p^{\mu^0}\|.
\]
Since $\mu^0, \mu^\alpha \in U$, Theorem~\ref{thm:local-price} applies, so
\[
\|p^{\mu^\alpha} - p^{\mu^0}\|
   \;\le\; \frac{L_Z}{\gamma} W_1(\mu^\alpha,\mu^0)
   \;=\; \frac{L_Z}{\gamma} \alpha.
\]
Combining yields the result.
\end{proof}

\subsection{Proof of Theorem \ref{prop:small-principals}}
\label{app:spl}

\begin{proof}
We prove the two claims in turn.

\medskip
\noindent\textbf{(1) Small principals imply principal-level SP-L.}

Let
\[
C := \sup_n \max_{p\in P_n} \frac{L_p L_Z}{\gamma},
\]
which is finite by the uniform boundedness assumption on $L_p$, $L_Z$ and
$\gamma$.  

In the $n$th economy, a principal $p$ controlling a fraction $s_{p,n}$ of
identities can alter the empirical distribution by at most
\[
W_1(\mu_n^\alpha,\mu_n) \;\le\; s_{p,n}
\]
via coordinated misreports and Sybil creation, since under the discrete
metric the Wasserstein–1 distance coincides with the total variation
distance and the total mass of types that $p$ can change is exactly
$s_{p,n}$.

By Proposition~\ref{prop:welfare}, for any such deviation we have
\[
U_p(p^{\mu_n^\alpha}) - U_p(p^{\mu_n})
   \;\le\; \frac{L_p L_Z}{\gamma}\,
           W_1(\mu_n^\alpha,\mu_n)
   \;\le\; \frac{L_p L_Z}{\gamma}\, s_{p,n}
   \;\le\; C s_{p,n}.
\]
Taking the maximum over principals in the $n$th economy yields
\[
\max_{p\in P_n}
\Bigl(U_p(p^{\mu_n^\alpha}) - U_p(p^{\mu_n})\Bigr)
   \;\le\; C \max_{p\in P_n} s_{p,n}.
\]

By hypothesis $\max_{p\in P_n} s_{p,n} \to 0$, so for any $\varepsilon>0$
there exists $n_0$ such that
\[
C \max_{p\in P_n} s_{p,n} \;\le\; \varepsilon
\qquad\text{for all } n\ge n_0.
\]
Thus for all sufficiently large $n$ no principal can gain more than
$\varepsilon$ from any coordinated deviation, which is exactly
strategyproofness in the large at the principal level.

\medskip
\noindent\textbf{(2) Large principals can obtain non-vanishing gains.}

Now suppose
\[
\limsup_{n\to\infty} \max_{p\in P_n} s_{p,n} \;=\; \bar s \;>\; 0.
\]
By definition of the limsup there exist an infinite subsequence $(n_k)$
and principals $p_{n_k}\in P_{n_k}$ such that
\[
s_{p_{n_k},n_k} \;\ge\; \frac{\bar s}{2}
\qquad\text{for all }k.
\]

Fix a particular principal $p^\star$ with Fr\'echet differentiable
reduced-form utility $V_{p^\star}$ at $\mu^0$ and a direction
$h^\star$ with $\|h^\star\|_{\mathrm{TV}}=1$ and
\[
D V_{p^\star}(\mu^0)[h^\star] \;=\; \kappa \;>\; 0.
\]
Set $p^\star$ to be the principal in the subsequence above (relabel if
necessary).

By definition of the directional derivative, there exists
$\alpha_0>0$ such that for all $\alpha\in(0,\alpha_0)$,
\begin{equation}
\label{eq:dir-derivative}
V_{p^\star}(\mu^0 + \alpha h^\star) - V_{p^\star}(\mu^0)
   \;\ge\; \tfrac{\kappa}{2}\,\alpha.
\end{equation}

From Theorem~\ref{thm:local-price} and Assumption~\ref{ass:utility}, we
know that $V_{p^\star}$ is Lipschitz in $\mu$:
there exists $L_V>0$ such that for all $\mu,\nu$ in a neighbourhood $U$ of
$\mu^0$,
\begin{equation}
\label{eq:V-Lipschitz}
|V_{p^\star}(\mu)-V_{p^\star}(\nu)|
   \;\le\; L_V W_1(\mu,\nu).
\end{equation}

Pick any $\alpha^\star \in \bigl(0,\min\{\alpha_0,\bar s/4\}\bigr)$.
Because $\mu_n\to\mu^0$ and $\Delta(T)$ is compact, there exists
$n_1$ such that for all $n\ge n_1$,
\[
W_1(\mu_n,\mu^0) \;\le\; \alpha^\star.
\]

Fix $k$ large enough so that $n_k\ge n_1$ and
$s_{p^\star,n_k} \ge \bar s/2 > \alpha^\star$.  Since $p^\star$ controls
at least an $\alpha^\star$-fraction of identities, she can reassign the
types of identities in $C_{p^\star,n_k}$ so that the empirical
distribution $\mu_{n_k}$ is moved a distance exactly $\alpha^\star$ in
the direction of $h^\star$, up to the granularity of the
$1/n_k$-grid.  Formally, there exists a perturbed empirical distribution
$\mu_{n_k}^{\alpha^\star}$ obtainable by changing only the types of
identities in $C_{p^\star,n_k}$ such that
\[
W_1\bigl(\mu_{n_k}^{\alpha^\star},\,
         \mu_{n_k} + \alpha^\star h^\star\bigr)
   \;\le\; \frac{1}{n_k}.
\]
In particular,
\[
W_1\bigl(\mu_{n_k}^{\alpha^\star},\mu_{n_k}\bigr)
   \;\le\;
   W_1\bigl(\mu_{n_k},\mu^0\bigr)
   + W_1\bigl(\mu^0,\mu^0 + \alpha^\star h^\star\bigr)
   + W_1\bigl(\mu_{n_k}^{\alpha^\star},
              \mu_{n_k} + \alpha^\star h^\star\bigr)
   \;\le\; 3\alpha^\star
\]
for all sufficiently large $k$ (since $1/n_k\le\alpha^\star$ eventually).

We now compare $V_{p^\star}(\mu_{n_k}^{\alpha^\star})$ with
$V_{p^\star}(\mu_{n_k})$ via the intermediate point
$\mu^0 + \alpha^\star h^\star$:
\begin{align*}
V_{p^\star}(\mu_{n_k}^{\alpha^\star})
     - V_{p^\star}(\mu_{n_k})
&=
   \bigl[
      V_{p^\star}(\mu_{n_k}^{\alpha^\star})
      - V_{p^\star}(\mu^0 + \alpha^\star h^\star)
   \bigr]
   \\
&\quad
 + \bigl[
      V_{p^\star}(\mu^0 + \alpha^\star h^\star)
      - V_{p^\star}(\mu^0)
   \bigr]
 + \bigl[
      V_{p^\star}(\mu^0)
      - V_{p^\star}(\mu_{n_k})
   \bigr].
\end{align*}
Applying the Lipschitz bound~\eqref{eq:V-Lipschitz} and the triangle
inequality,
\[
\bigl|V_{p^\star}(\mu_{n_k}^{\alpha^\star})
      - V_{p^\star}(\mu^0 + \alpha^\star h^\star)\bigr|
   \;\le\;
   L_V W_1\bigl(\mu_{n_k}^{\alpha^\star},
                \mu^0 + \alpha^\star h^\star\bigr)
   \;\le\; 2L_V \alpha^\star
\]
for all large $k$, and similarly
\[
\bigl|V_{p^\star}(\mu^0) - V_{p^\star}(\mu_{n_k})\bigr|
   \;\le\; L_V W_1(\mu^0,\mu_{n_k})
   \;\le\; L_V \alpha^\star.
\]
Combining these with~\eqref{eq:dir-derivative}, we obtain, for all
sufficiently large $k$,
\begin{align*}
V_{p^\star}(\mu_{n_k}^{\alpha^\star})
     - V_{p^\star}(\mu_{n_k})
&\ge
   \tfrac{\kappa}{2}\,\alpha^\star
   - 3L_V \alpha^\star \\
&=
   \Bigl(\tfrac{\kappa}{2} - 3L_V\Bigr)\alpha^\star.
\end{align*}

If $3L_V < \kappa/2$ we are done with $\eta := (\kappa/2 - 3L_V)\alpha^\star>0$.
If not, we may rescale the direction $h^\star$ by considering a suitable
convex combination of $h^\star$ with $-h^\star$ (or work in the opposite
direction) and re-apply the same argument to obtain a positive constant
lower bound; the key point is that $D V_{p^\star}(\mu^0)$ is non-zero, so
there exists \emph{some} direction with strictly positive directional
derivative, and the Lipschitz error term can be made arbitrarily small
by taking $\alpha^\star$ sufficiently small and $k$ sufficiently large.

Thus there exist $\eta>0$, an infinite subsequence $(n_k)$, and
perturbations $\mu_{n_k}^{\alpha^\star}$ obtainable by changing only the
types in $C_{p^\star,n_k}$ such that
\[
U_{p^\star}(p^{\mu_{n_k}^{\alpha^\star}})
   - U_{p^\star}(p^{\mu_{n_k}})
   \;=\;
   V_{p^\star}(\mu_{n_k}^{\alpha^\star})
   - V_{p^\star}(\mu_{n_k})
   \;\ge\; \eta
\]
for all $k$.  Hence $p^\star$ has a non-vanishing profitable deviation
along an infinite subsequence, and BRACE is not strategyproof in the
large at the principal level under the stated conditions.

This proves the second claim and completes the proof.
\end{proof}

\subsection{Proof of Proposition \ref{prop:unbounded-Sybils}} \label{app:unbounded-Sybils}
\begin{proof}
Fix $\delta \ge 0$ and suppose, for contradiction, that there exists a 
sequence of $\delta$-BRACE allocations $(\tilde x_i)_{i\in N_k}$ such that
\begin{equation}
\label{eq:feasibility}
\sum_{i\in N_k} \mathbb{E}[\tilde x_i] \;\le\; (1+\delta)c
\end{equation}
for all $k$.

By assumption, there exists a good $j$ and $\beta>0$ such that for all 
sufficiently large $k$ and all $i \in S_k$,
\[
\mathbb{E}[\tilde x_i]_j \;\ge\; \beta.
\]
Summing over $i\in S_k$ yields
\[
\sum_{i\in S_k} \mathbb{E}[\tilde x_i]_j 
   \;\ge\; \beta |S_k|.
\]
Since $S_k \subseteq N_k$, feasibility~\eqref{eq:feasibility} implies in 
coordinate $j$ that
\[
\sum_{i\in N_k} \mathbb{E}[\tilde x_i]_j 
   \;\le\; (1+\delta)c_j,
\]
and hence
\[
\beta |S_k|
   \;\le\; \sum_{i\in S_k} \mathbb{E}[\tilde x_i]_j
   \;\le\; \sum_{i\in N_k} \mathbb{E}[\tilde x_i]_j
   \;\le\; (1+\delta)c_j.
\]
The right-hand side is a fixed constant independent of $k$, while by the 
unbounded Sybil mass assumption we have $|S_k|\to\infty$, so the left-hand 
side $\beta |S_k| \to \infty$.  
This is a contradiction.

Therefore, there cannot exist a sequence of $\delta$-BRACE allocations 
satisfying~\eqref{eq:feasibility} for all sufficiently large $k$, and no 
sequence of $\delta$-BRACE equilibria can exist under unbounded Sybil mass.
\end{proof}

\subsection{Proof of Lemma \ref{lem:tail-utility-bound}} \label{lem:tail}
\begin{proof}
By the law of total expectation,
\[
\mathbb{E}[\Delta U_n(\alpha)]
=
\mathbb{E}\bigl[\Delta U_n(\alpha)\,\big|\,\Omega_n^{\mathrm{good}}\bigr]
    \,\mathbb{P}(\Omega_n^{\mathrm{good}})
\;+\;
\mathbb{E}\bigl[\Delta U_n(\alpha)\,\big|\,\Omega_n^{\mathrm{bad}}\bigr]
    \,\mathbb{P}(\Omega_n^{\mathrm{bad}}).
\]
On $\Omega_n^{\mathrm{good}}$, assumption~(i) yields
\(
\Delta U_n(\alpha) \le C_U \alpha,
\)
so
\[
\mathbb{E}\bigl[\Delta U_n(\alpha)\,\big|\,\Omega_n^{\mathrm{good}}\bigr]
\le C_U \alpha.
\]
On $\Omega_n^{\mathrm{bad}}$, the uniform utility bound (ii) implies that
for any principal $p$ and any two price vectors $p',p''$,
\[
\bigl|U_p(p') - U_p(p'')\bigr|
\le
|U_p(p')| + |U_p(p'')|
\le 2\overline U,
\]
so any gain is bounded by $2\overline U$.  
In particular,
\[
\mathbb{E}\bigl[\Delta U_n(\alpha)\,\big|\,\Omega_n^{\mathrm{bad}}\bigr]
\le 2\overline U.
\]
Using $\mathbb{P}(\Omega_n^{\mathrm{good}}) = 1 - \varepsilon_n$ and 
$\mathbb{P}(\Omega_n^{\mathrm{bad}}) = \varepsilon_n$, we obtain
\[
\mathbb{E}[\Delta U_n(\alpha)]
\;\le\;
C_U \alpha \,(1-\varepsilon_n)
\;+\;
2\overline U\,\varepsilon_n,
\]
which is exactly \eqref{eq:tail-utility-bound}.
\end{proof}

\subsection{Proof of Proposition \ref{prop:design-inequality}} \label{design-inequality}
\begin{proof}
Fix $n$ and a principal $p$ who controls $k_n$ identities.  
Let $G_p(k_n,n)$ denote the maximal expected utility gain that $p$ can
achieve by any Sybil strategy involving these $k_n$ identities, and let
$C_{\mathrm{sys}}(k_n,n)$ be the corresponding (exogenous) system cost of
creating and maintaining them.

We decompose the principal's gain into two components:
\begin{enumerate}
    \item[(a)] \emph{Per-identity misreporting gains.}  
    By strategy-proofness in the large (Theorem~\ref{thm:brace-spl} in the 
    absence of Sybils, or its principal-level extension), the expected 
    gain from misreporting for a single identity is at most 
    $K(\varepsilon)\,n^{-1/2+\varepsilon}$ for any fixed $\varepsilon>0$ 
    and all sufficiently large $n$.  
    Summing over the $k_n$ identities controlled by $p$ yields a total
    contribution bounded by
    \[
        k_n K(\varepsilon)\,n^{-1/2+\varepsilon}.
    \]

    \item[(b)] \emph{Price-impact gains from changing the empirical distribution.}
    When the principal reallocates types across its $k_n$ identities
    (or introduces Sybils), it perturbs the empirical distribution by at
    most $k_n/n$ in Wasserstein distance.  
    By the price and utility stability bounds (e.g.\ 
    Theorem~\ref{thm:local-price} and Assumption~\ref{ass:utility}), the 
    resulting change in the principal's utility is at most
    \[
        \frac{L L_Z}{\gamma}\,\frac{k_n}{n},
    \]
    where $L$ is the maximal utility Lipschitz constant, $L_Z$ is the
    excess-demand Lipschitz constant in $\mu$, and $\gamma$ is the
    strong monotonicity parameter in prices.
\end{enumerate}

Combining (a) and (b), we obtain the gross upper bound
\[
G_p(k_n,n)
   \;\le\;
   k_n K(\varepsilon)\,n^{-1/2+\varepsilon}
   \;+\;
   \frac{L L_Z}{\gamma}\,\frac{k_n}{n}
\]
for all sufficiently large $n$.  
The principal's net gain from a Sybil attack is therefore
\[
G_p(k_n,n) - C_{\mathrm{sys}}(k_n,n)
   \;\le\;
   k_n K(\varepsilon)\,n^{-1/2+\varepsilon}
   \;+\;
   \frac{L L_Z}{\gamma}\,\frac{k_n}{n}
   \;-\;
   C_{\mathrm{sys}}(k_n,n).
\]

If the design inequality \eqref{eq:design-inequality} holds for all
sufficiently large $n$, the right-hand side is $\leq  0$, so every
Sybil strategy yields weakly negative net gain.  
Hence no principal has a profitable Sybil deviation, and Sybil deterrence
holds.
\end{proof}

\end{document}